\documentclass[twocolumn,amsmath,amssymb]{revtex4}
\usepackage{graphicx}
\usepackage{dcolumn}
\usepackage{bm}
\begin{document}

\
\title{Proposed Graphene Nanospaser}

\author{Vadym Apalkov$^1$}
\author{Mark I. Stockman$^{1,2,3}$}
\affiliation{
$^1$Department of Physics and Astronomy, Georgia State
University, Atlanta, Georgia 30303, USA\\
$^2$Fakult\"at f\"ur Physik, Ludwig-Maximilians-Universit\"at, Geschwister-Scholl-Platz 1, D-80539 M\"unchen, Germany\\
$^3$Max-Planck-Institut f\"ur Quantenoptik, Hans-Kopfermann-Strasse 1, D-85748 Garching, Germany
}

\date{\today}
%


\maketitle 

{\bf 
Spaser (Surface Plasmon Amplification by Stimulated
Emission of Radiation) \cite{Bergman_Stockman:2003_PRL_spaser, Stockman_Nat_Phot_2008_Spasers_Explained, Stockman_JOPT_2010_Spaser_Nanoamplifier,
Noginov_et_al_Nature_2009_Spaser_Observation, Hill_et_al_Opt_Expr_2009_Polaritonic_Nanolaser, Oulton_Sorger_Zentgraf_Ma_Gladden_Dai_Bartal_Zhang_Nature_2009_Nanolaser, 
Gwo_et_al_Science_2012_Spaser}, which is a nanoplasmonic counterpart of laser, consists of a plasmonic 
nanoparticle coupled to an active (gain) medium. Spasing  results in coherent generation of surface plasmons with nanolocalized, intense, coherent optical fields. One of the limitations on the spaser is posed by a high damping rate of the surface plasmon excitations in plasmonic metals.
Here, we propose a novel nanospaser in the mid-infrared utilizing a nanopatch of graphene  as its plasmonic core and a quantum-well cascade as its gain medium. This design takes advantage of low optical losses in graphene due to its high electron mobility  \cite{Novoselov_Geim_et_al_nature04233_2D_Electrons_in_Graphene, Kim_et_al_2005_nature04235_Quantum_Hall_Effect_in_Graphene}. In contrast with earlier designs of quantum cascade lasers using plasmonic waveguides \cite{Capasso_et_al_APL_1995_QCL_Plasmonic_Waveguide, Capasso_et_al_APL_2000_QCL_with_SPP_Waveguide} and subwavelength resonators \cite{Faist_et_al_Science_2010_THz_microlaser},
the proposed quantum cascade graphene spaser generates optical fields with unprecedented high nanolocalization, which is characteristic of  
graphene plasmons \cite{Abajo_et_al_PRL_2013_e187401_Giant_Enhencement_in_Graphene_Dimers}. 
This spaser will be an efficient nanosource of intense and coherent hot-spot fields in the mid-infrared spectral region with wide perspective applications in mid-infrared nanoscopy, nanospectroscopy,  nanolithography, and optoelectronic information processing.
}

Quantum nanoplasmonics is becoming a fast-growing field of nanooptics \cite{Stockman_Opt_Expres_2011_Nanoplasmonics_Review} due to unique property of plasmonic 
system to enhance strongly subwavelength optical fields while maintaining their quantum properties. The plasmonic
enhancement is a basis of many applications of nanoplasmonics\cite{Moskovits:1985_RMP_SERS, 
Kall_et_al_PRL_1999_SERS_Hemoglobin_Molecule, Tang_et_al_Nat_Phot_2008_Dipole_Nanoantenna_Ge_Photodetector,  Nagatani_et_al_Sci_Techn_Adv_Mater_2006_Immunochromatographic_Tests}.
There is especial interest in mid-infrared near-field spectroscopy because of distinct spectral signatures of important molecular, biological, and condensed-matter systems, including macromolecules and graphene \cite{Hillenbrand_et_al_APL_2005_Near_IR_SNOM, Basov_et_al_Nature_2012_Graphene_mid_IR_Plasmonics}. For further development of plasmonics and near-field optics and their applications, it is crucial to have a coherent source of intense local fields on the nanoscale, in particular, in the mid-infrared.

\begin{figure}
\begin{center}\includegraphics[width=0.4\textwidth]{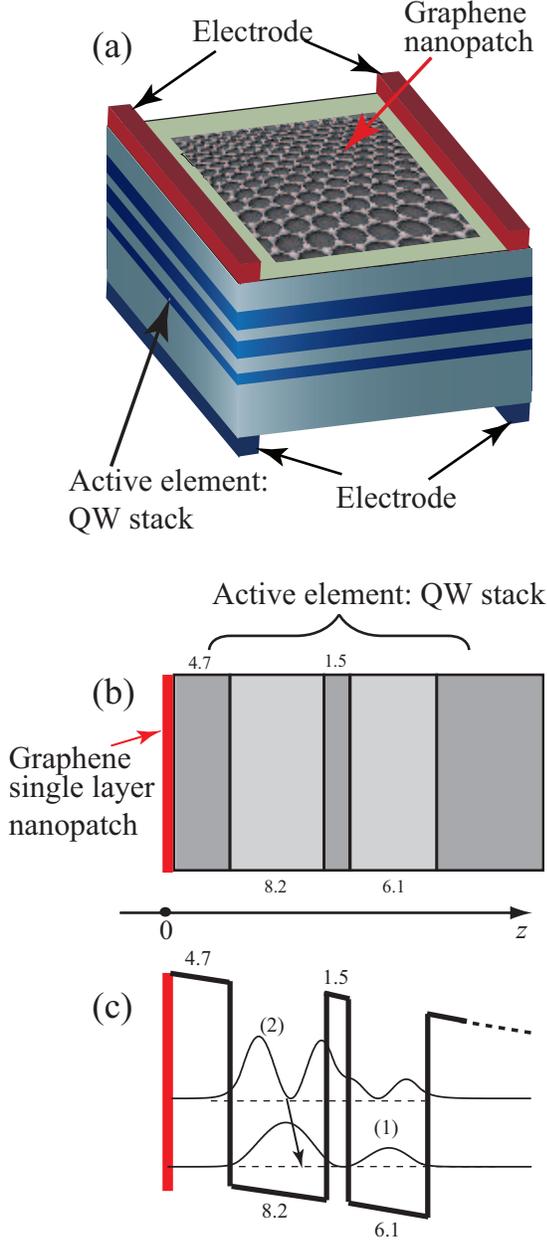}\end{center}
\caption{Schematics of graphene quantum cascade nanospaser. (a) Geometry of the spaser: monolayer graphene  nanopatch atop of a stack of quantum-cascade wells and electrodes for electric pumping (the graphene may be used as one of the electrodes). (b) Schematics of cross section geometry of the quantum cascade nanospaser. The numbers (in nm) are the widths of the wells ($\mathrm{Ga}_{0.47}\mathrm{In}_{0.53}\mathrm{As}$) and 
barriers ($\mathrm{Al}_{0.48}\mathrm{In}_{0.52}\mathrm{As}$). These wells and barriers are 
shown by light and dark gray colors, respectively. (c) Energy-band diagram of the active region of a 
quantum cascade laser structure under electric field bias. Relative wave functions of subbands (1) 
and (2) as well as the optical transition corresponding to the spaser action are shown schematically.  
}
\label{Fig_graphene}
\end{figure}

Here, we propose  a novel spaser as  
coherent quantum generator of surface plasmons in nanostructured graphene. The plasmonic core of this spaser is a graphene-monolayer nanopatch, and its active (gain) element is a
cascade quantum well system with a design similar to the design of an active element of quantum cascade lasers -- see Fig.\ \ref{Fig_graphene} (a).

\begin{figure}
\begin{center}\includegraphics[width=0.4\textwidth]{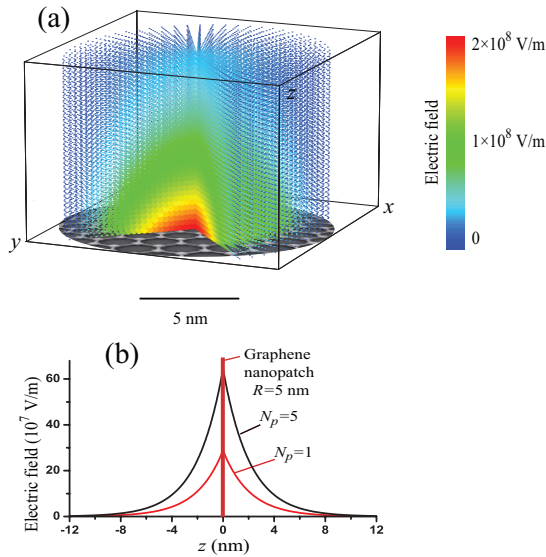}\end{center}
\caption{Optical electric field distribution in graphene nanospaser. (a) Spatial distribution of local optical electric field generated in a spaser with a single surface plasmon in the  ground mode on a circular graphene nanopatch of $R=5$ nm radius. The three-dimensional  distribution (shown over the graphene nanodisk only) is displayed as a "field cloud" where the field  is depicted by points whose color and size correspond to the local amplitude. The spatial scale and the field color-coded scale for the field are indicated. One quarter of the distribution is cut off to better display the inner structure of the field. Schematic of the graphene nanopatch is also shown at the bottom. (b) Vertical (normal to graphene nanopatch surface) amplitude of the electric field is plotted as a function of $z$ at the center of the graphene nanodisk for two values of the the plasmon population: $N_p=1$ and $N_p=5$ as indicated.
}
\label{Graphene_Spaser_Fields.pdf}
\end{figure}

Graphene is a layer of crystalline carbon that is just one atom thick 
\cite{Novoselov_nat_mater_2007, Electronic_properties_graphene_RMP_2009, graphene_advances_2010}. Its
low-energy electronic excitations are described by the
chiral relativistic massless Dirac-Weyl equation. Doped graphene is known to posses plasmonic properties  with a potentially high plasmonic quality factor (low optical loss) due to high electron mobility \cite{Novoselov_Geim_et_al_nature04233_2D_Electrons_in_Graphene, Kim_et_al_2005_nature04235_Quantum_Hall_Effect_in_Graphene, Abajo_et_al_Nano_Lett_2011_Graphene_Plasmonics, plasmon_infrared_frequencie_PRB_2011,
Grigorenko_et_al_Nat_Phot_2012_Graphene_Plasmonics, Basov_et_al_Nature_2012_Graphene_mid_IR_Plasmonics}. The size of the graphene nanopatch may be as small as $\lesssim 10$ nm. Graphene possesses unprecedented tight localization of local plasmonic fields at the surface, which provides an unique opportunity to build an electrically-pumped mid- to near-infrared nanospaser with coherent optical fields of extremely high concentration and intensity -- see Fig.\ \ref{Graphene_Spaser_Fields.pdf}. This will be the first electrically pumped nanospaser: the previous nanospasers were optically pumped \cite{Noginov_et_al_Nature_2009_Spaser_Observation, Gwo_et_al_Science_2012_Spaser} while the electrically pumped spasers had mode volumes of microscale size at least in one dimension \cite{Hill_et_al_Opt_Expr_2009_Polaritonic_Nanolaser, Oulton_Sorger_Zentgraf_Ma_Gladden_Dai_Bartal_Zhang_Nature_2009_Nanolaser}.

The chiral nature of the electronic states in graphene results in 
strong suppression of electron scattering, which greatly increases the mobility 
 of charge carriers \cite{Novoselov_Geim_et_al_nature04233_2D_Electrons_in_Graphene} and greatly 
reduces the theoretically-estimated plasmonic losses \cite{Abajo_et_al_Nano_Lett_2011_Graphene_Plasmonics}. For plasmon frequency $\omega_q$ below the interband loss threshold, $2E_F/\hbar$, where  $E_F$ is the Fermi energy of graphene, and below the optical phonon frequency, $\hbar\omega_o\approx 0.2~\mathrm{eV}$, the plasmon relaxation time can be estimated from direct-current measurements.
In unsuspended graphene, the high static carrier mobility of $\mu \approx 2.5\times 10^4~\mathrm{cm}^2
\mathrm{V}^{-1}\mathrm{s}^{-1}$ \cite{Kim_et_al_2005_nature04235_Quantum_Hall_Effect_in_Graphene, plasmon_infrared_frequencie_PRB_2011}
 is realized at carrier density $n\approx 5\times 10^{12}$ 
$\mathrm{cm}^{-2}$, which corresponds to $E_F =\hbar v_F \sqrt{\pi n}\approx 0.3$ eV, where $v_F\approx 1.15\times 10^8~\mathrm{cm/s}$ is Fermi velocity \cite{Tutuc_et_al_PRL_2012_Graphene_Fermi_Energy}.
For such mobility, the plasmon relaxation time in graphene can be estimated as $\tau = \mu \hbar \sqrt{\pi n}/ev_F\approx 0.6$ ps. This yields plasmon quality factor
$Q=\omega_q\tau/2 \approx 50 $ at frequency $\hbar \omega_q = 130$ meV.

As the active element of the graphene spaser, we consider a multi-quantum well system, which is similar to the design of the active element of quantum cascade laser \cite{kazarinov_QCL_1971, Capasso_QCL_Science_1994}.
 Within the active region of such a multi-quantum well structure,  optical transitions occur between subband levels of dimensional quantization in the growth direction.  

In the graphene spaser, we utilize only one period (a single multi-quantum well heterostructure) from the active medium of the quantum cascade laser -- see Fig.\ \ref{Fig_graphene} where one of  possible designs of the active medium is illustrated. The functioning of such an active medium is characterized by fast and efficient electrical injection of electrons into the upper subband level of the multi-quantum well system and their  extraction from the lower subband level -- see  Fig.\ \ref{Fig_graphene} (c). Since the plasmonic modes of graphene are strongly localized in the normal direction (i.e., in the growth direction of the quantum wells, along the $z$ axis in Fig.\ \ref{Fig_graphene}), with localization length $\sim 3-5$ nm, only one period of the quantum cascade gain medium couples to the plasmonic modes of graphene. The gain, i.e., population inversion, in the quantum well system is introduced through a voltage applied to the system.

To find plasmonic modes of graphene, we assume that the graphene monolayer is surrounded by two homogeneous media with dielectric constants $\epsilon_1 $ and $\epsilon_2$. Effectively, this approximation amounts to replacement of the cascade system with a homogeneous effective medium.We assume that graphene monolayer is positioned in a plane $z=0$. Its plasmon wave function (electric field) in the upper half plane ($z>0$) is $\propto e^{iqy} e^{-k_{1z} z}$ and in the lower half plane  ($z<0$) is $\propto e^{iqy} e^{k_{2z}z}$, where $q$ is the in-plane wave vector and $k_{1z}$ and $k_{2z}$ are the decay constants in the upper and lower half planes, respectively.  

 In the Drude approximation, we obtain [see Eq. (8) of Supplementary Information (SI)] the well-known dispersion relation in graphene for a TM plasmon mode with frequency $\omega_q$ and wave vector $q$,
\begin{equation}
q \approx k_{1z} \approx k_{2z} \approx
 \frac{ \hbar^2  (\epsilon_1 + \epsilon_2)}{4e_0^2 E_F}\omega_q \left( \omega_q + i\gamma\right)  ~,
\label{omega0}
\end{equation}
where  electron polarization relaxation rate $\gamma $  determines plasmon dephasing time $\tau=1/\gamma$. 

In the quantum description of the graphene spaser, we introduce the Hamiltonian of the spaser in the form
\begin{eqnarray}
{\cal H} & = &  \sum_{i=1,2} \sum _{\mathbf{k}}  E_i(k) \hat{c}_{i,\mathbf{k}} ^\dag \hat{c}_{i,\mathbf{k}} 
 + \hbar \sum _\mathbf{q} \omega_q \hat{a}^\dag_\mathbf{q} \hat{a}_\mathbf{q}   \nonumber \\
 & & + \hbar \sum _{\mathbf{q},\mathbf{k}} \Omega _{12}(k,q) \left( \hat{a}^\dag_\mathbf{q} \hat{c}_{1,\mathbf{k}} ^\dag \hat{c}_{2,\mathbf{k}} + h.c. \right)~,
 \label{H1}
\end{eqnarray}
where the first sum describes the Hamiltonian of the gain medium, which consists of two subbands $i=1$ and $i=2$ (see Fig.\ \ref{Fig_graphene}) with energies 
$E_1(k) = \varepsilon_1 +\hbar ^2 k^2/2m^\ast$ and 
$E_2(k) = \varepsilon_2 +\hbar ^2 k^2/2m^\ast$, respectively,   
$\varepsilon_1 <\varepsilon_2$. Here $m^\ast$ is electron effective mass,  and $\mathbf k$ is 
a two-dimensional electron wave vector. The second sum in Eq.\ (\ref{H1}) is the Hamiltonian of plasmons in the graphene layer, where  $\hat{a}_q$ and $\hat{a}^\dag_q$ are operators of annihilation and creation of a plasmon with wave vector $q$  [see Eq.\ (9) of SI]. The last sum is the interaction Hamiltonian, $-\hat{\mathbf{E}} \hat{\mathbf {d}}$, of plasmonic field $\hat{\mathbf{E}}$ with dipole $\hat{\mathbf{d}}$ of the gain medium. 

Quantization of graphene surface plasmons yields modes  [see Eqs. (1)-(2) and (11) of SI] that possess  extremely  tight nano-confinement at the graphene nanopatch surface as illustrated in Fig.\ \ref{Graphene_Spaser_Fields.pdf}. These local fields form a hot spot shown by the red color in  Fig.\ \ref{Graphene_Spaser_Fields.pdf} (a), which has the size of only a few nanometers in each of the three dimensions. These fields are coherent and their amplitude in this hot spot is very high, ${\cal E}\sim 100~\mathrm{MV/m}$ for one plasmon per mode; the field amplitude increases with the number of plasmons as ${\cal E}\propto N_p^{1/2}$. The unprecedented field localization characteristic of graphene and their high amplitudes inherent in nanospasers bear potential for various important applications, from nanospectroscopy and sensing to nanolithography.

The dipole interaction strength is determined by the corresponding Rabi frequency, $\Omega_{12}$, 
\begin{equation}
\Omega_{12}(\mathbf{k},\mathbf{q}) = e_0 {\cal E}_q \int d\mathbf{r} \psi^\ast_{1,\mathbf{k}} (\mathbf{r}) E_{q,z} z \psi_{2,\mathbf{k}}(\mathbf{r}),
\label{rabi1}
\end{equation} 
where we have taken into account that for intersubband transitions the dipole matrix element has only the $z$ component, $\psi_{1,k}$ and $\psi_{2,k}$ are electron wave functions with wave vector $\mathbf k$ of subbands 1 and 2, respectively. 
 
Assuming for simplicity that electron dynamics in the $x$, $y$, and $z$ directions is separable, we can factorize the wave functions,
$\psi_{i,\mathbf k}(\mathbf r) =  S_0^{-1/2}\chi_i (z)  \exp (i\mathbf{k} \mathbf{r}_\perp)$, where $\mathbf{r}_\perp = (x,y)$ is a two-dimensional radius. Then Eq.\ (\ref{rabi1}) becomes 
\begin{equation}
\Omega_{12}(\mathbf{k},\mathbf{q}) =  e_0{\cal E}_q \int dz \chi^\ast_{1}(z) e^{-k_{2z} z} z \chi_{2}(z) \approx {\cal E}_q  d_{12} e^{-qL},
\label{Omega12}
\end{equation} 
where $d_{12} =  e_0\int dz \chi^{*}_{1} (z) z \chi_{2} (z) $ is the intersubband envelope dipole matrix element, and $L$ is the distance between the graphene monolayer and the active quantum well of the cascade structure. 

Introducing the density matrix of the active quantum well of the gain medium, $\rho _{\mathbf{k}} (t)$, we describe the dynamics of the spaser within the density matrix approach, writing the equation for the density matrix, $i\hbar 
\dot{\rho} _{\mathbf{k}} (t) = [ \rho _{\mathbf{k}} (t), {\cal H}]$. For the stationary regime, we obtain the condition of spasing in the form \cite{Bergman_Stockman:2003_PRL_spaser, Stockman_JOPT_2010_Spaser_Nanoamplifier} 
\begin{equation}
\frac{(\Gamma_{12}+\gamma)^2}{(\omega_{12} - \omega_q)^2 + (\Gamma _{12} + \gamma )^2}
\sum _{\mathbf{k}} \left| \Omega _{12}(\mathbf{k},\mathbf{q}) \right|^2 \geq \gamma \Gamma_{12},
\label{spasing1}
\end{equation}
where $\Gamma_{12}$ is the intersubband polarization-relaxation rate, and $\omega_{12} = (\varepsilon _2 - \varepsilon_1)/\hbar$ is the intersubband transition frequency.
From this and Eq.\ (\ref{Omega12}), substituting ${\cal E}_q $ from Eq.\ (12) of SI, we obtain 
\begin{equation}
\frac{(\gamma + \Gamma_{12})^2}{(\omega_{12} - \omega_q)^2 + (\Gamma _{12} + \gamma )^2}
\frac{\hbar \omega_q q k_0^2 }{ (\epsilon_1 + \epsilon_2)} |d_{12}|^2 e^{-2qL} 
\geq \gamma \Gamma_{12}.
\label{spasing2}
\end{equation}
Here wave vector $k_0$ is an effective Fermi wave vector, i.e., the largest wave vector of the states, which are partially occupied by electrons in the gain region. Only such states contribute to the sum in Eq.\ (\ref{spasing1}). 

At the resonance, i.e. at $\omega_{12} = \omega_q$, the condition of spasing simplifies to 
\begin{equation}
\frac{\hbar \omega_q q k_0^2 }{\epsilon_1 + \epsilon_2} |d_{12}|^2 e^{-2qL} 
\geq \gamma \Gamma_{12}.
\label{spasing3}
\end{equation}
For given active region parameters, Eq.\ (\ref{spasing3}) introduces a constraint on plasmon polarization-relaxation rate $\gamma$. 

Consider realistic parameters of the active region [corresponding, e.g., 
to the structure shown in Fig.\ \ref{Fig_graphene} (b)]: $\hbar \omega_{12} = 
130 $ meV, $d_{12} = 1.5~ e_0\mathrm{nm}$, and $L\approx 16 $ nm. A typical intersubband relaxation 
time in the quantum well cascade systems is $\tau _{12} = \hbar /\Gamma_{12} \approx 1$ ps \cite{Bastard_relaxation_PRB_89, QCL_linewidth_theory}. Under the resonant condition,
the intersubband transition frequency $\hbar \omega_{12} = 130 $ meV, which is equal to the plasmon 
frequency, corresponds to the plasmon wave vector  $q \approx 
0.3 $  nm$^{-1}$ -- see Eq.\ (\ref{omega0}). Realistically assuming that $k_0 \approx 0.04$ nm$^{-1}$ and $\epsilon_1 + \epsilon_2 \approx 12$, we obtain 
\begin{equation}
\gamma \leq 2\cdot 10^{-3} \mathrm{eV}
\end{equation}
with the corresponding relaxation time $\tau \geq 3.6 \cdot 10^{-13}$ s. 

\begin{figure}
\begin{center}\includegraphics[width=0.4\textwidth]{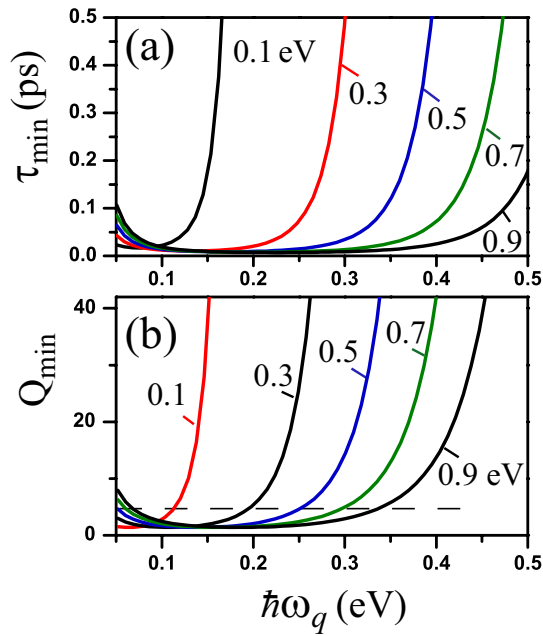}\end{center}
\caption{(a) Minimum relaxation time of plasmons in graphene sufficient for spasing, as defined by Eq.\ (\ref{spasingTAU}),  is shown as a 
function of plasmon frequency for different Fermi energies of electrons in graphene. The spasing occurs for plasmon relaxation time $\tau > \tau _{\mathrm{min}}$. 
(b) Minimum plasmon quality factor $Q_{\mathrm{min}}$, which is sufficient for spasing, as defined 
by Eq.\ (\ref{spasingGamma}), is shown as a function of plasmon frequency $\omega_q$. The dashed line 
shows the experimental value of plasmon quality factor  $Q \approx 4$ \cite{Basov_et_al_Nature_2012_Graphene_mid_IR_Plasmonics}.
The numbers next to the lines in pane;s (a) and (b) are the Fermi energies in eV. 
}
\label{Fig_tau_min}
\end{figure}

Substituting relation (\ref{omega0}) into inequality (\ref{spasing3}), we obtain a spasing condition as a limitation on the plasmon polarization-relaxation time as
\begin{equation}
\tau \geq \tau_{\mathrm{min}} = \frac{ 4 e_0^2 \Gamma_{12} E_F  }{ \hbar^2 \omega_q^3 k_0^2 |d_{12}|^2} e^{2qL}.
\label{spasingTAU}
\end{equation}
The dependence of the minimum plasmon relaxation time, $\tau_{\mathrm{min}}$, on the plasmon 
frequency $\omega _q$ is shown in Fig.\ \ref{Fig_tau_min}(a) for different Fermi energies $E_F$ of electrons 
in graphene. 
The data clearly show that there is a  broad range of plasmon frequencies in the mid-infrared $\hbar \omega _q \approx 
0.1 - 0.3 $ eV where the required minimum relaxation time is relatively short, $\tau_{\mathrm{min}}\sim 0.02$ ps, so the spasing threshold condition (\ref{spasingTAU}) can realistically be satisfied.
Outside this range, the minimum relaxation time greatly increases. Fortuitously, typical values of frequencies 
of intersubband transitions in quantum wells are within the same mid-infrared range 
$0.1 - 0.3 $ eV. 

The spasing condition of Eq.\ (\ref{spasingTAU}) can be also expressed in terms of dimensionless parameter -- plasmon quality factor $Q =\omega_q \tau/2=\omega_q /(2\gamma)$,
\begin{equation}
Q\ge Q_\mathrm{min}~.
\label{spasingGamma}
\end{equation}
Condition (\ref{spasingGamma}) on the surface plasmon quality factor depends in the plasmon frequency, $\omega_q$, which under resonant condition is equal to the frequency of  intersubband 
optical transitions in the active region of the cascade system. As a function of frequency $\omega_q$, 
the minimum sufficient quality factor, $Q_{\mathrm{min}}$, has a deep minimum, the position of which is determined by 
condition $q\approx 1/2L$ or $\omega_q \approx \left( 2e_0^2 E_F/[L\hbar^2 (\epsilon_1 + \epsilon_2) ]\right)^{1/2}$. 
 It depends on parameters of the system, 
such as the Fermi energy of electrons in graphene, $E_F$, and the distance, $L$, between the graphene monolayer and the active quantum well of the cascade structure. 
The dependence of  $Q_{\mathrm{min}}$ on 
the frequency of surface plasmons, $\omega_q$, is shown in Fig.\ \ref{Fig_tau_min}(b) 
for different values of Fermi energy $E_F$. The minimum value of $Q$ is 
$Q_{\mathrm{min}}=\left. 2e_0 \Gamma_{12} (\epsilon_1 + \epsilon_2)\right/\left[k_0^2 |d_{12}|^2\right]\approx 0.1$, 
and it does not depend on the Fermi energy, which affects only the position of this minimum. This minimum required quality factor of the plasmons is very low, indeed, which implies that the spasing is relatively easy to achieve.
In fact, the quality factor value, $Q\approx 50$, which is obtained for graphene from experimental data as discussed at the beginning in the introductory part of this Letter,  is high enough, $Q\gg Q_{\mathrm{min}}$ (it is on the same order as for silver in the visible region\cite{Stockman_Opt_Expres_2011_Nanoplasmonics_Review}),  for spasing to occur in a broad  range of frequencies $\omega_q$ -- see \ Fig.\ \ref{Fig_tau_min} (b).


Plasmon damping rate in graphene was measured experimentally by mid-infrared nanoscopy  \cite{Basov_et_al_Nature_2012_Graphene_mid_IR_Plasmonics}.  
The best fit to the experimental data was obtained for rather strong damping or a low quality factor, $Q\approx 4$, at plasmon frequency 
$\hbar \omega_q = 0.11$ eV, which corresponds to  plasmon relaxation time $\tau = 0.044$ ps. 
Even for such strong plasmon damping, the plasmon quality factor, $Q\approx 4$, is much greater than the minimum required value 
rate  $Q_{\mathrm{min}}\sim 0.1$  -- see Fig.\ \ref{Fig_tau_min}(b).
Taking into account that the plasmon relaxation rate strongly increases above the  optical phonon frequency, $\hbar\omega_o\approx 0.2$ eV, we can conclude that the optimal  Fermi energy in graphene should be  $E_F\approx0.15$ eV. 
For such Fermi energy, the minimum of $Q_{\mathrm{min}}$ is realized at the plasmon frequency in the mid-infrared,
$\hbar \omega_q \approx 0.15$ eV -- see Fig.\ \ref{Fig_tau_min}(b). 
The above results clearly suggests that the spasing can be realized in 
realistic graphene systems coupled to quantum cascade well structures. 

In conclusion, we predict that a graphene monolayer nanopatch deposited on a top of a specially designed stack of quantum wells 
constitutes an electrically-pumped spaser, which is a coherent source of surface plasmons and  intense local optical fields in graphene in the mid-infrared. The system of quantum wells
is designed similar to an active element of quantum cascade laser and is characterized by efficient electrical injection of electrons into an upper subband level of multi-quantum well system and fast 
extraction of electrons from a lower subband level. Under realistic parameters of the multiple quantum wells, our analysis shows that the condition of spasing, i.e., coherent generation of plasmons in graphene, can be achieved under realistic conditions at 
finite doping of graphene monolayer corresponding to electron Fermi energy in graphene $\sim 0.15$ eV. 
The above analysis can be applied not only to a graphene monolayer, but also for graphene-based 
nanoscale structures such as graphene nanoribbons, graphene quantum dots, and graphene nanoantennas, in particular, nanodimers where a possibility of efficient nanofocusing has been recently proposed \cite{Abajo_et_al_PRL_2013_e187401_Giant_Enhencement_in_Graphene_Dimers}. The predicted graphene spaser will be an efficient source of intense and coherent nanolocalized fields in the mid-infrared spectrum  -- cf.\ Fig.\ \ref{Graphene_Spaser_Fields.pdf}. With  size of such a near-field localized mid-infrared source $\lesssim 10$ nm being on the same  order of magnitude as important biological objects (viruses, cell organelles, chromosomes, large proteins or DNA) and technological elements (transistors and interconnects), such a spaser will have 
wide perspective applications in mid-infrared nanoscopy, nanospectroscopy,  nanolithography, biomedicine, nano-optoelectronics, etc.  


\section*{Acknowledgments}

This work has been supported by Grant No. DE-FG02-11ER46789 from the Materials Sciences and Engineering Division and by Grant No. DE-FG02-01ER15213 from the Chemical Sciences, Biosciences and Geosciences Division, Office of the Basic Energy Sciences, Office of Science, U.S. Department of Energy. 
MIS acknowledges support by the Max Planck Society and the Deutsche Forschungsgemeinschaft Cluster of Excellence: Munich Center for Advanced Photonics (http://www.munich-photonics.de).

\section*{Competing financial interests}

The authors declare no competing financial interests.

\end{document}